\documentclass{article}
\usepackage{amssymb}
\usepackage{amsmath,bm}
\usepackage{array}

\def\un{{\rm 1\mkern-4mu I}}
\begin{document}

\begin{flushright}
CERN-PH-TH/2010-071
\end{flushright}
\vskip 1in
\begin{center}
{\Large Minimal representations and reductive dual pairs in}
\vskip 0.2cm
{\Large conformal field theory}\footnote{Invited lecture at the 8th
International Workshop on Lie Theory and Its Applications in Physics, Varna,
June 2009; ed. V. Dobrev, AIP Conference Proceedings {\bf 1243} (Melville, NY 
2010) pp. 13-30.}
\vskip 0.5cm
{\large Ivan Todorov}\footnote{Permanent address: Institute for Nuclear Research
 and Nuclear Energy, Bulgarian Academy of Sciences, Tsarigradsko Chaussee 72,
 BG-1784 Sofia, Bulgaria;
e-mail: todorov@inrne.bas.bg}
\vskip 0.5cm
\small Theory Division, Department of Physics, CERN
\\ CH-1211 Geneva 23, Switzerland
\end{center}


\vglue 2cm

\begin{abstract}
A {\it minimal representation} of a simple non-compact Lie group is obtained by ``quantizing'' the minimal nilpotent coadjoint orbit of its Lie algebra. It provides context for Roger Howe's notion of a reductive dual pair encountered recently in the description of global gauge symmetry of a (4-dimensional) conformal observable algebra. We give a pedagogical introduction to these notions and  point out that physicists have been using both minimal representations and dual pairs without naming them and hence stand a chance to understand their theory and to profit from it.
\end{abstract}

\newpage

\tableofcontents

\bigskip

\noindent {\bf References}

\newpage

\section{Introduction}\label{sec1}

\begin{itemize}
\item[ \ ] {\sl ``$\ldots$ for me the motivation is mostly the desire to understand the hidden machinary in a striking concrete example, around which one can build formalisms''.}\footnote{M. Kontsevich, Beyond numbers, in: {\it The Unravelers, Mathematical Snapshots}, ed. by J.-F.~Dars, A.~Lesne, A.~Papillaut, A.K.~Peters, Wellesley, MA 2008, p.~182.}
\end{itemize}

\bigskip

The concept of a minimal representation, introduced in the mid 1970's by A.~
Joseph \cite{J74} \cite{J76} (triggered by the physicists' excitement with
spectrum generating algebras) keeps attracting attention of mathematicians and
(mathematical) physicists (for a review up to the late 1990's -- see
\cite{Li00}; for subsequent developments --  see e.g. \cite{G01,G06} \cite{K02}
 \cite{GS05} \cite{KM} \cite{F09} \cite{Li09} \cite{FG09} \cite{K10} and
references therein). We shall review the identification of the minimal orbit --
 the classical counterpart of the minimal representation -- in Section~2. We
will also display, in Section~3, an old physical example -- the
{\it oscillator} (called in the 1960's {\it ladder}) {\it representations} of
$U(2,2)$ which describe zero-mass particles \cite{MT69}(constructed before the
concept of a minimal representation was coined).

\smallskip

The second item of our survey, the notion of a {\it (reductive) dual pair}, was introduced by Roger Howe in an influential preprint of the 1970's that was eventually published as \cite{H89}. It was previewed in a lecture \cite{H85} highlightening its applications to physics. The principle underlying such applications was, in fact, anticipated by the founders of the axiomatic approach to quantum field theory who introduced the concept of superselection sectors \cite{W$^3$} and developed it systematically in the work of Doplicher, Haag and Roberts (for a review and further references -- see \cite{H}). In this framework the algebra of observables and the global (compact) gauge group are mutual commutants in the (extended) field algebra, thus providing a generic example of a dual pair. A more elementary example is provided by a pair of subgroups, $G_1 , G_2$, of a group $G$, each being the centralizer of the other in $G$. One can then ask how does an irreducible representation (IR) of $G$ split into IRs of the pair $(G_1 , G_2)$. This question is, in general, not easy to answer. It becomes manageable if the given IR of $G$ is a minimal one. We review in Section~4 the concept of a reductive dual pair and display Howe's classification of such {\it irreducible pairs} acting by automorphisms on a {\it complex symplectic space}.

\smallskip

In Section~5 we prepare the stage for an extension and application of the above notions to an infinite dimensional framework: {\it globally conformal invariant quantum field theory} in four spacetime dimensions. The assumption of global conformal invariance (GCI) \cite{NT} allows to extend the concept of {\it vertex algera} -- the mathematicians' way of viewing the chiral observable algebra of {\it $2$-dimensional conformal field theory} (2D~CFT) -- to any higher even dimensional theory \cite{N}. We also introduce there the notion of a {\it bilocal field} which opens the way to use infinite dimensional Lie algebras in the context of 4D~CFT. In Section~6 we review earlier work \cite{B07,B08} which demonstrates that the physical (unitary, positive energy) representations of the observable algebra ${\mathcal L}$ is a Fock space representation that splits into IRs whose multiplicities are dictated by a compact gauge group $U$, the pair $({\mathcal L} , U)$ appearing as a dual pair in the field algebra extension of ${\mathcal L}$ envisaged by Haag et al. \cite{H}. Before concluding we also review in Section~6 the basic properties and the oscillator representations of the non-compact orthogonal group $SO^* (4n)$ (combining material of \cite{BB} and of \cite{B08}).

\section{Minimal coadjoint orbits}\label{sec2}

A coadjoint orbit of a Lie group $G$ is an orbit of the coadjoint representation $Ad'_g$ of $G$ acting on the vector space ${\mathcal G}'$ dual to the Lie algebra ${\mathcal G}$ of $G$. Each coadjoint orbit ${\mathcal O}$ admits a $G$-invariant symplectic form and hence is an (even dimensional) symplectic manifold which can be interpreted as the phase space of a  classical Hamiltonian system. The {\it orbit method} reflects the expectation that the unitary IRs of $G$ can be obtained by ``quantizing'' such classical symplectic orbits. It has been introduced by A.A.~Kirillov in 1962 and further developed (along with the {\it geometric quantization}, chiefly) by Kirillov, Kostant, Duflo, and Souriau -- see for a readable introduction and references \cite{K99}.

\smallskip

We recall some basic notions and facts of the theory of {\it semisimple Lie algebras} ${\mathcal G}$. An {\it element} $s \in {\mathcal G}$ is called {\it semisimple} if it is conjugate to an element of a Cartan subalgebra of ${\mathcal G}$ (or, equivalently, if $ad_s$ is diagonalizable). An $n \in {\mathcal G}$ is said to be {\it nilpotent} if $ad_n$ is nilpotent (i.e., if a finite power of $ad_n$ vanishes).

\bigskip

\noindent {\bf Fact.}  Every element $x \in {\mathcal G}$ has a unique representation of the form $x = s+n$ where $s$ is semisimple, $n$ is nilpotent, and $[s,n]=0$. The non-degenerate Killing form $B(X,Y)$ on ${\mathcal G}$ allows to define an isomorphism of ${\mathcal G}$-spaces ${\mathcal G} \leftrightarrow {\mathcal G}'$. The invariance of $B$ implies that there exists a bijection between adjoint and coadjoint orbits that preserves semi-simplicity and nilpotency. (We say that a coadjoint orbit ${\mathcal O}$ is semisimple/nilpotent if every (equivalently some) element of ${\mathcal O}$ has this property.)

\smallskip

A complex semisimple Lie algebra ${\mathcal G}_{\mathbb C}$ has a unique (up to
 conjugation) Cartan subalgebra ${\mathfrak h}$. The set of semisimple coadjoint orbits of ${\mathcal G}_{\mathbb C}$ is isomorphic to ${\mathfrak h} / W$ where $W$ is the Weyl group of ${\mathcal G}_{\mathbb C}$; hence, there is a continuum of such orbits. By contrast, the (partially ordered) set ${\mathcal N}$ of {\it nilpotent orbits} is finite (see \cite{CMcG} for details and proofs -- also of Proposition~2.1 below -- and Section~5 of \cite{F09} for a summary).

\bigskip

\noindent {\bf Proposition 2.1.} --  (a) {\it There exists a unique principal nilpotent orbit ${\mathcal O}_{\rm princ}$ of maximal dimension
\begin{equation}
\label{eq2.1}
\vert {\mathcal O}_{\rm princ} \vert = \vert {\mathcal G} \vert - {\rm rank} \, {\mathcal G} \, ,
\end{equation}
 where by $\vert {\mathcal M} \vert$ we denote the dimension of the manifold ${\mathcal M}$.}

 \medskip

\noindent (b) {\it There is a unique subregular nilpotent orbit, ${\mathcal O}_{\rm subreg}$, of dimension
 \begin{equation}
\label{eq2.2}
\vert {\mathcal O}_{\rm subreg} \vert = \vert {\mathcal G} \vert - {\rm rank} \, {\mathcal G} - 2 \, .
\end{equation}
}

\medskip

\noindent (c) {\it There exists a unique non-zero nilpotent orbit ${\mathcal O}_{\rm min}$ of minimal dimension.}

\bigskip

\noindent {\bf Remark 2.1.} The principal orbit is dense in ${\mathcal N}$ with respect to the {\it Zariski topology}. Its closure consists of all lower dimensional nilpotent orbits. The subregular orbit is Zariski dense in ${\mathcal N} \backslash {\mathcal O}_{\rm princ}$. The {\it Zariski closure} of ${\mathcal O}_{\rm min}$ is ${\mathcal O}_{\rm min} \cup \{ 0 \}$.

\bigskip

\noindent {\bf Example.} The semisimple orbits of $s\ell_2$ are
\begin{equation}
\label{eq2.3}
{\mathcal O}_k = \left\{ \begin{pmatrix} x &y \\ z &-x \end{pmatrix} , \ x,y,z \in {\mathbb C} \, ; \ x^2 + yz = k \ne 0 \right\} \, .
\end{equation}
For $k=0$, the corresponding ${\mathcal O}_0$ is the unique non-trivial nilpotent orbit: ${\mathcal O}_0 = {\mathcal O}_{\rm princ} = {\mathcal O}_{\rm min}$ (${\mathcal O}_{\rm subreg}$ is trivial in this case -- consisting of the $0$-element of $s\ell_2$).

\bigskip

To construct the minimal orbit of a simple Lie algebra ${\mathcal G}$ we observe that all simple complex Lie algebras have an essentially unique 5-grading (as exploited in \cite{G01,K02,G06}):
\begin{equation}
\label{eq2.4}
{\mathcal G} = {\mathcal G}_{-2} \oplus {\mathcal G}_{-1} \oplus {\mathcal G}_0 \oplus {\mathcal G}_1 \oplus {\mathcal G}_2 \, ,
\end{equation}
given by the eigenvalues of the Cartan generator $H_{\theta}$, corresponding to the highest root $\theta$. Indeed the relations
\begin{equation}
\label{eq2.5}
[H_{\theta} , E_{\alpha}] = 2 \, \frac{(\alpha \mid \theta)}{(\theta \mid \theta)} \, E_{\alpha} \, , \quad 2 \, \frac{(\alpha \mid \theta)}{(\theta \mid \theta)} \in {\mathbb Z} \, , \quad \vert (\alpha \mid \theta) \vert \leq ( \theta \vert \theta)
\end{equation}
imply that the coefficient to  $E_{\alpha}$ in the right hand side belongs to the set $\{ -2 , -1 , 0,1,2 \}$. The spaces ${\mathcal G}_{\pm 2}$ have each dimension 1 and are generated by $E_{\pm \theta}$ satisfying $[E_{\theta} , E_{-\theta}] = H_{\theta}$. The triple $\{ E_{\theta} , H_{\theta} , E_{-\theta} \}$ thus spans a distinguished subalgebra $s\ell_2$ of ${\mathcal G}$. Let ${\mathcal H}$ be the maximal Lie subalgebra of ${\mathcal G}$ commuting with this principal $s\ell_2$. The minimal orbit in the complexification ${\mathcal G}_{\mathbb C}$ of ${\mathcal G}$, isomorphic to the adjoint orbit of $E_{\theta}$, has dimension
\begin{equation}
\label{eq2.6}
\vert {\mathcal O}_{\rm min} \vert = \vert {\mathcal G}_1 \vert + 2 =: 2 \dim \pi_{\theta}
\end{equation}
where $\pi_{\theta}$ is the corresponding {\it minimal representation} and $\dim \pi_{\theta}$ is its {\it Gelfand-Kirillov dimension} -- roughly speaking, the number of variables on which depends the wave function of the associated Schr\"odinger picture \cite{KM}. Knowing ${\mathcal G}$ and ${\mathcal H}$ we can compute $\vert {\mathcal G}_1 \vert$ and hence $\dim \pi_{\theta}$
\begin{equation}
\label{eq2.7}
\vert {\mathcal G} \vert = \vert {\mathcal H} \vert + 2 \, \vert {\mathcal G}_1 \vert + 3 \, .
\end{equation}
In Table~1 we list the type of ${\mathcal H}$ and the values of $\vert {\mathcal G}_1 \vert = \frac{1}{2} \, (\vert {\mathcal G} \vert - \vert {\mathcal H} \vert - 3)$ and of $\dim \pi_{\theta}$ for all complex simple Lie algebras (cf. \cite{G06}).
$$
\begin{tabular}{|c|c|c|c|c|}
${\mathcal G}_{\mathbb C}$ &$\vert {\mathcal G} \vert$ &${\mathcal H}$ &$\vert {\mathcal G}_1 \vert \! = \! \frac{1}{2} (\vert {\mathcal G} \vert \! - \! \vert {\mathcal H} \vert \! - \! 3)$ &$\dim_{\mathbb C} \pi_{\theta}$ \\ \hline
$s\ell_n = A_{n-1}$ &$n^2 -1$ &$g\ell_{n - 2}$ &$2(n-2)$ &$n-1$ \\
$so(2n+1)=B_n$ &$n(2n+1)$ &$B_{n-2} \oplus s\ell_2$ &$2(2n-3)$ &$2n-2$ \\
$sp(2n)=C_n$ &$n(2n+1)$ &$C_{n-1}$ &$2(n-1)$ &$n$ \\
$so(2n)=D_n$ &$n(2n-1)$ &$D_{n-2} \oplus s\ell_2$ &$4(n-2)$ &$2n-3$ \\
$E_6$ &78 &$s\ell_6$ &20 &11 \\
$E_7$ &133 &$D_6$ &32 &17 \\
$E_8$ &248 &$E_7$ &56 &29 \\
$F_4$ &52 &$C_3$ &14 &8 \\
$G_2$ &14 &$s\ell_2$ &4 &3 \\
\end{tabular}
$$

\centerline{{\bf Table 1.} -- {\it List of ${\mathcal G}_{\mathbb C} \supset s\ell_2 \oplus {\mathcal H}$;  dimensions.}}

\bigskip

The next step, the ``quantization of ${\mathcal O}_{\rm min}$'', is not canonical. It has been best studied for $sp(2n,{\mathbb R})$: it is realized in the familiar Fock space (or oscillator) representation of the Heisenberg algebra (of $n$ creation and $n$ annihilation operators) that will be reviewed in the next section. It is also referred to as the Segal-Shale-Weil representation, made popular by the work of Weil \cite{W} with its applications to $\theta$-functions and automorphic forms -- an important chapter in number theory (for a systematic pedagogical review of this case -- see \cite{F}. The case of $O(p,q)$ is developed more recently and surveyed in the monograph \cite{KM} and in the recent review
article \cite{K10}.)

\section{Examples: $u(n,n) \subset sp(4n,{\mathbb R})$. Massless particles}\label{sec3}
\setcounter{equation}{0}

We begin, following Howe \cite{H85}, with the Heisenberg algebra of $n$ annihilation and $n$ creation operators $c = (c_i , i=1,\ldots , n)$ and $c^* = (c_j^* ; j = 1 , \ldots , n)$ satisfying the {\it canonical commutation relations} (CCR)
\begin{equation}
\label{eq3.1}
[c_i , c_j] = 0 = [c_i^* , c_j^*] \, , \quad [c_i , c_j^*] = \delta_{ij} \, , \quad i,j = 1 , \ldots , n \, .
\end{equation}
The CCR are left invariant by the $n(2n+1)$ dimensional real symplectic group $Sp(2n,{\mathbb R})$ whose elements $g$ can be written in terms of a pair of complex $n \times n$ matrices $U$ and $V$:
\begin{equation}
\label{eq3.2}
g = \begin{pmatrix} U &\bar V \\ V &\bar U \end{pmatrix} \, , \quad gc = Uc + \bar V c^* \, , \quad gc^* = Vc + \bar U c^* \, ,
\end{equation}
where $\bar U$ and $\bar V$ are complex conjugate to $U$ and $V$ and we have
\begin{equation}
\label{eq3.3}
UV^* = \bar V \, ^tU \, , \quad UU^* - \bar V \, ^tV = \un
\end{equation}
($U^*$ is the hermitean conjugate to $U$, $U^* = \, ^t\bar U$, $^tA$ standing for the transposed of the matrix $A$). We note that the pair $(c,c^*)$ provides a complex variable parametrization of a real $2n$ dimensional vector space; a similar remark applies to the above parametrization of the real symplectic group $Sp (2n,{\mathbb R})$.

\smallskip

A more symmetric way of writing the CCR consists in introducing a $2n$-component vector $\xi = (\xi_a , a=1 ,\ldots , 2n)$ and a {\it symplectic form} $J = (J_{ab})$ such that
\begin{eqnarray}
\label{eq3.4}
\lbrack \xi_a , \xi_b \rbrack &= &J_{ab} \quad ( = -J_{ba}) \nonumber \\
JJ^* &= &-J^2 = \un \quad (= (\delta_{ab})) \, .
\end{eqnarray}
The invariance of the CCR (\ref{eq3.4}) under symplectic transformations then reads
\begin{equation}
\label{eq3.5}
g \, J \ ^tg = J \qquad \mbox{for} \quad g \in Sp (2n,{\mathbb R}) \, .
\end{equation}
The relations (\ref{eq3.1}) correspond to the choice $\xi = (c,c^*)$, $J = \begin{pmatrix} {\mathbb O} &\un \\ -\un &{\mathbb O} \end{pmatrix}$.

\smallskip

The {\it oscillator representation} of the Lie algebra $sp(2n,{\mathbb R})$ is spanned by the anticommutators $\frac{1}{2} \, (\xi_a \, \xi_b + \xi_b \, \xi_a)$. The {\it Fock space representation} of the Heisenberg algebra (with {\it ket} and {\it bra} vacuum vectors $\vert 0 \rangle$ and $\langle 0 \vert$ such that $c_i \, \vert 0 \rangle = 0 = \langle 0 \vert \, c_i^*$, $i = 1 , \ldots , n$; $\langle 0 \mid 0 \rangle = 1$) splits into two most degenerate IRs of $sp(2n,{\mathbb R})$. They can be only exponentiated to unitary IRs of the double cover of $Sp (2n,{\mathbb R})$, the {\it metaplectic group} $Mp (2n)$ (that is not a matrix group). The resulting Fock space representation of $Mp(2n)$ corresponds to a minimal coadjoint orbit, thus providing an example of a minimal representation. We shall exploit this in the physically interesting case of $n=4$, displaying on the way the reduction of the minimal representation of $Mp(8)$ into the zero mass representations of the (helicity extended) spinorial conformal group $U(2,2)$ (reviewing the construction of \cite{MT69}).

\smallskip

To make explicit the inclusion $u(2,2) \subset sp (8,{\mathbb R})$ we shall label the annihilation (and the creation) operators by $a_1 , a_2 , b_1 , b_2$ (and $a_1^* , \ldots , b_2^*$). With this notation the $u(2,2)$ sub (Lie) algebra of $sp (8,{\mathbb R})$ is defined as the centralizer of (twice) the helicity operator,
\begin{equation}
\label{eq3.6}
h := a_1^* \, a_1 + a_2^* \, a_2 -  b_1^* \, b_1 - b_2^* \, b_2 \, .
\end{equation}
If we introduce the (Dirac conjugate, operator valued) spinors
\begin{eqnarray}
\label{eq3.7}
\varphi = \begin{pmatrix} a_1 \\ a_2 \\ b_1^* \\ b_2^* \end{pmatrix} , &&\tilde\varphi = \varphi^* \beta = (a_1^* , a_2^* , -b_1 , -b_2) \ (\beta = {\rm diag} (1,1,-1,-1)) \, , \nonumber \\
&& [\varphi^{\alpha} , \tilde\varphi_{\alpha'}] = \delta_{\alpha'}^{\alpha} \ (\alpha , \alpha' = 1,2,3,4)
\end{eqnarray}
then the oscillator realization of $u(2,2)$ is given by operators of the form $\hat X = \tilde\varphi \, X \, \varphi$ where $X$ are $4 \times 4$ matrices satisfying the antihermiticity condition $X^* \beta + \beta X = 0$. A Chevalley-Cartan basis of $su(2,2)$ is given by
$$
E_1 = a_1^* \, a_2 \, , \ E_2 = a_2^* \, b_1^* \, , \ E_3 = -b_1 \, b_2^* \, , \ F_1 = a_2^* \, a_1 \, , F_2 = -b_1 \, a_2 \, , \ F_3 = -b_2 \, b_1^* \, ,
$$
\begin{equation}
\label{eq3.8}
H_i = [E_i , F_i] \, , \ H_1 = a_1^* \, a_1 - a_2^* \, a_2 \, , \ H_2 = a_2^* \, a_2 + b_1 \, b_1^* \, , \ H_3 = b_2^* \, b_2 - b_1^* \, b_1
\end{equation}
reproducing the standard Cartan matrix $(c_{ij})$ for $s\ell_4 : [H_i , E_j] = c_{ij} \, E_j$). The positive roots correspond to the raising operators $E_i$, $[E_1 , E_2] = a_1^* \, b_1^*$, $[E_2 , E_3] = a_2^* \, b_2^*$ and the highest root
\begin{equation}
\label{eq3.9}
E_{\theta} = [[E_1 , E_2],E_3] = a_1^* \, b_2^* \, .
\end{equation}
The two other generators of the distinguished $s\ell_2$ ($=su(1,1)$) are
\begin{equation}
\label{eq3.10}
(E_{-\theta} \equiv) \ F_{\theta} = -b_2 \, a_1 \, , \ H_{\theta} = [E_{\theta} , F_{\theta}] = a_1^* \, a_1 + b_2 \, b_2^* \ ([H_{\theta} , E_{\theta}] = 2E_{\theta}) \, .
\end{equation}
Its commutant ${\mathcal H}$ in $su(2,2)$ is the 4-dimensional $u(1,1)$ subalgebra spanned by products of $(a_2^* , b_1)$ with $(a_2 , b_1^*)$.

\bigskip

\noindent {\bf Remark 3.1.} Splitting, more generally, the annihilation (creation) operators
$c_1^{(*)} , \ldots , c_{2n}^{(*)}$ into ``positive and negative charge'' ones, $a_1^{(*)} , \ldots , a_n^{(*)} , b_1^{(*)} , \ldots , b_n^{(*)}$ we can realize, similarly, the embedding $u(n,n) \subset sp (4n)$ where $u(n,n)$ is the centralizer of the {\it charge operator}
$$
Q := a_1^* \, a_1 + \ldots + a_n^* \, a_n - b_1^* \, b_1 - \ldots - b_n^* \, b_n \, .
$$
The non-compact Chevalley generators are then $E_n = a_n^* \, b_1^*$, $F_n = -b_1 \, a_n$ while the highest root operator becomes $E_{\theta} = a_1^* \, b_n^*$. The commutant ${\mathcal H}$ of the distinguished $s\ell_2$, spannned by $E_{\theta}$, $F_{\theta} = -b_n \, a_1$, $H_{\theta} = a_1^* \, a_1 + b_n \, b_n^*$, is $u(n-1 , n-1)$ and the minimal coadjoint orbit is $4n-2$ dimensional.

\bigskip

Returning to the case $n=2$, the 6-dimensional minimal coadjoint orbit is spanned by $E_{\theta} , H_{\theta}$ and by the elements $E_1 , E_3 , [E_1 , E_2]$ and $[E_2 , E_3]$ of ${\mathcal G}_1$ (of (\ref{eq2.3})). We shall introduce light-cone ``momentum space'' variables that give rise to a Schr\"odinger like realization of the minimal representation (in the state space of massless particles). To this end we introduce two complex conjugate pairs of variables $z_{\alpha}$ and $\bar z_{\alpha}$, $\alpha = 1,2$, setting:
\begin{eqnarray}
\label{eq3.11}
&&a_{\alpha}^* = \frac{1}{\sqrt 2} \, (z_{\alpha} - \bar\partial_{\alpha}) \, , \ b_{\alpha}^* = \frac{1}{\sqrt 2} \, (\bar z_{\alpha} - \partial_{\alpha}) \, , \ \partial_{\alpha} := \frac{\partial}{\partial \, z_{\alpha}} \, , \ \bar\partial_{\alpha} := \frac{\partial}{\partial \, \bar z_{\alpha}} \, , \nonumber \\
&&a_{\alpha} = \frac{1}{\sqrt 2} \, (\bar z_{\alpha} + \partial_{\alpha}) \, , \ b_{\alpha} = \frac{1}{\sqrt 2} \, (z_{\alpha} + \bar\partial_{\alpha})
\end{eqnarray}
or
$$
z_{\alpha} = \frac{1}{\sqrt 2} \, (a_{\alpha}^* + b_{\alpha}) \, , \ \bar z_{\alpha} = \frac{1}{\sqrt 2} \, (a_{\alpha} + b_{\alpha}^*) \, , \ \partial_{\alpha} = \frac{1}{\sqrt 2} \, (a_{\alpha} - b_{\alpha}^*) \, , \ \bar\partial_{\alpha} = \frac{1}{\sqrt 2} \, (b_{\alpha} - a_{\alpha}^*) \, .
$$
Then the 4-vector
\begin{equation}
\label{eq3.12}
p_{\mu} = z \, \sigma_{\mu} \, \bar z \quad \mu = 0,1,2,3 \ \left( \sigma_0 = \begin{pmatrix} 1 &0 \\ 0 &1 \end{pmatrix} \right)
\end{equation}
($\sigma_1 , \sigma_2 , \sigma_3$ being the Pauli matrices) is light-like $(p_0 = \vert {\bm p} \vert)$ and has been proved to correspond to the translation generators of the Poincar\'e subalgebra of $su(2,2)$ -- i.e. to the 4-momentum (see \cite{MT69}). Thus, our representation indeed corresponds to $0$-mass particles and can be realized in a space of functions of three real variables ${\bm p} = (p_1 , p_2 , p_3)$, in accord with the value $\dim \pi_{\theta} = 3$ ($= \frac{1}{2} \, \vert {\mathcal O}_{\rm min} \vert$) of its Gelfand-Kirillov dimension. The irreducible components of this representation are labeled by the (integer) eigenvalues of the operator $h = \frac{1}{2} \, (\tilde\varphi \, \varphi + \varphi \, \tilde\varphi)$ (\ref{eq3.4}) (which, according to the conventions of \cite{MT69}, are equal to minus twice the helicity). The vacuum vector belongs to the zero helicity IR and has the form
$$
\vert 0 \rangle = \frac{2}{\pi} \, e^{-z\bar z} \, \left( = \frac{2}{\pi} \, e^{-p_0} \right) \, , \ z \, \bar z = z_1 \, \bar z_1 + z_2 \, \bar z_2 \, ,
$$
\begin{equation}
\label{eq3.13}
\langle 0 \mid 0 \rangle = \frac{4}{\pi^2} \int d^2 \, z_1 \int d^2 \, z_2 \, e^{-2z\bar z} = 4 \int_0^{\infty} p_0 \, e^{-2 p_0} \, dp_0 = 1 \, .
\end{equation}
It can be defined as the unique lowest weight vector (satisfying, by definition, $F_i \, \vert 0 \rangle = 0$, $i=1,2,3$) that transforms under a 1-dimensional representation of the maximal compact subgroup $S(U(2) \times U(2))$ of $SU(2,2)$:
$$
E_1 \, \vert 0 \rangle = 0 = H_1 \, \vert 0 \rangle = E_3 \, \vert 0 \rangle = H_3 \, \vert 0 \rangle \, ,
$$
\begin{equation}
\label{eq3.14}
(H_1 + 2 \, H_2 + H_3) \, \vert 0 \rangle = (z \, \bar z - \bar\partial \, \partial) \, \frac{2}{\pi} \, e^{-z\bar z} = 2 \, \vert 0 \rangle
\end{equation}
($H_1 + 2 \, H_2 + H_3$ being the generator of the centre, $u(1)$, of $s(u(2) \oplus u(2))$). The vacuum thus plays the role of a {\it spherical vector}.

\bigskip

The Fock space (${\mathcal F}$-) representation of the Heisenberg algebra of $(a_{\alpha}^{(*)} , b_{\beta}^{(*)})$ splits into two IRs of $sp(8,{\mathbb R})$ which, in turn, split into IRs of $u(2,2)$ involving either, only even or only odd eigenvalues of $h$ (cf. \cite{I67}). The minimal coadjoint orbit of $Mp(8,{\mathbb R})$, the orbit of the highest root $\frac{1}{2} \, a_1^{*2}$, is 8 dimensional (see Table~1), so that the associated Gelfand-Kirillov dimension is four (the 4th quantum number being the helicity).

\smallskip

The subgroup $U(2,2)$ and its centre $U(1)$ provide an example of a reductive dual pair in $Mp(8)$, $U(1)$ playing the role of a (global) gauge group.

\bigskip

\noindent {\bf Remark 3.2.} The subspace of ${\mathcal F}$ spanned by functions  $e^{-z\bar z} \, P(z_1 , z_2)$ where $P$ is a polynomial provides a model for the IRs of $SU(2)$, its Hilbert space closure being isomorphic to the Bargmann space of analytic functions (see \cite{B}).

\section{Two types of reductive dual pairs}\label{sec4}
\setcounter{equation}{0}

Let us look more closely at the example of $U(1)$, $U(2,2)$ as a dual pair in $Mp(8)$. The symplectic group $Sp(8,{\mathbb R})$ acts by symplectomorphisms on the 8-dimensional vector space ${\mathcal W}$ spanned by $(a_1 , a_2 , b_1 , b_2 ; a_1^* , a_2^* , b_1^* , b_2^*)$: it preserves both their commutation relations and their conjugation properties. (So does its double cover $Mp(8)$ which, however, does not act faithfully on ${\mathcal W}$.) The ``4-spinors'' $\varphi$ and $\tilde\varphi$ (\ref{eq3.7}) define a (complex) {\it Lagrangian polarization} in ${\mathcal W}$:
\begin{equation}
\label{eq4.1}
{\mathcal W} = {\mathcal V} \oplus {\mathcal V}' \, , \quad \varphi^{\alpha} \in {\mathcal V} \, , \quad \tilde\varphi_{\alpha'} \in {\mathcal V}'
\end{equation}
as $[\varphi^{\alpha} , \varphi^{\beta}] = 0 = [\tilde\varphi_{\alpha'} , \tilde\varphi_{\beta'}]$, while the CCR $[\varphi^{\alpha} , \tilde\varphi_{\alpha'}] = \delta_{\alpha'}^{\alpha}$ allow to regard the elements of ${\mathcal V}'$ as linear functionals on ${\mathcal V}$; thus, the subspaces ${\mathcal V}$ and ${\mathcal V}'$ of ${\mathcal W}$ are dual to each other. The subgroup $U(2,2)$ of $Mp(8)$, viewed as an automorphism group of ${\mathcal W}$, preserves separately the subspaces ${\mathcal V}$ and ${\mathcal V}'$ -- i.e., the Lagrangian polarization. Sure, $U(1) \subset U(2,2)$ (the subgroup of automorphisms $u_t (\varphi) = e^{-it} \, \varphi$, $u_t (\tilde\varphi) = e^{it} \, \tilde\varphi$, generated by the helicity invariant $h$ (\ref{eq3.6})) also preserves both ${\mathcal V}$ and ${\mathcal V}'$.

\smallskip

A simple example of a (reductive) dual pair which does not preserve any Lagrangian subspace of ${\mathcal W}$ is given by the entire group of symplectomorphism ($Sp(8)$, in our example) and its centre which changes the sign of all creation and annihilation operators. The 2-element group ${\mathbb Z} / (2)$ of the centre of $Sp(2n)$ can be viewed as the 1-dimensional orthogonal group $O(1)$. It turns out that our two examples are simple prototypes of a general classification which we proceed to describe.

\smallskip

Let $G$ and $G'$ be a dual pair of subgroups of $Mp(2n)$. Then they act by symplectomorphisms on the $2n$ dimensional space ${\mathcal W}$. They are said to be {\it reductive} iff for each $G^{(')}$-invariant subspace ${\mathcal V} \subset {\mathcal W}$ its complement $\tilde{\mathcal V}$ (such that ${\mathcal W} = {\mathcal V} \oplus \tilde{\mathcal V}$) is also $G^{(')}$ invariant. We shall only deal with reductive dual pairs and will skip the qualification ``reductive''. If ${\mathcal W}_1$ and ${\mathcal W}_2$ are two symplectic vector spaces and if $(G_1 , G'_1) , (G_2 , G'_2)$ are two dual pairs in the corresponding metaplectic groups, then the direct products ($G = G_1 \times G_2$, $G' = G'_1 \times G'_2$) will form a dual pair for the symplectic vector space ${\mathcal W} = {\mathcal W}_1 \oplus {\mathcal W}_2$. Such a dual pair is called {\it reducible}. A dual pair $(G,G')$ acting on ${\mathcal W}$ is called {\it irreducible} if it cannot be represented that way. An irreducible dual pair $(G,G')$ acting on ${\mathcal W}$ is said to be of {\it type II} if both $G$ and $G'$ leave a Lagrangian subspace of ${\mathcal W}$ invariant and of {\it type I} otherwise. These two types exhaust all possibilities of irreducible dual pairs acting on a {\it complex} symplectic space ${\mathcal W}$ (or, more generally, if the ground field is algebraically closed). For the case of a real ${\mathcal W}$ a classification appears in the second reference \cite{H89}. We shall content ourselves by giving some generic examples that generalize those discussed in the beginning of this section.

\bigskip

We first consider the tensor product ${\mathcal W}_{2mn} := {\mathbb R}^m \otimes {\mathcal W}_{2n}$ where ${\mathcal W}_{2n}$ is a (real) $2n$ dimensional space with symplectic form $J = (J_{ab})$ (\ref{eq3.4}). Viewing ${\mathbb R}^m$ as an orthogonal vector space with a $O(m)$-invariant scalar product we can equip ${\mathcal W}_{2mn}$ with a symplectic form setting in an appropriate basis $\xi_{ia}$, $i = 1,\ldots , m$, $a=1,\ldots , 2n$
\begin{equation}
\label{eq4.2}
[\xi_{ia} , \xi_{jb}] = \delta_{ij} \, J_{ab} \, .
\end{equation}
Then $(O(m) , Mp(2n))$ form a dual pair of type I in $Mp(2mn)$ (our $O(1)$ example above appearing as a special case).

\smallskip

To generalize our type II example, $(U(1),U(2,2))$, we consider the tensor product space
\begin{equation}
\label{eq4.3}
{\mathcal W}_{8n} := {\mathbb C}^n \otimes ({\mathcal V} \oplus {\mathcal V}')
\end{equation}
where ${\mathcal V}$ and ${\mathcal V}'$ are the dual spaces of Dirac conjugate spinors $\varphi$ and $\tilde\varphi$ (see \ref{eq4.1})). Regarding ${\mathbb C}^n$ as a unitary space equipped with a $U(n)$ invariant (positive) hermitean form we can introduce a symplectic structure on ${\mathcal W}_{8n}$ by defining (in a suitable basis $\varphi_i^{\alpha}$, $\tilde\varphi_{j\beta}$) the non-trivial CCR as
\begin{equation}
\label{eq4.4}
[\varphi_i^{\alpha} , \tilde\varphi_{j\beta}] = \delta_{ij} \, \delta_{\beta}^{\alpha} \, , \quad i,j = 1,\ldots , n \, ; \ \alpha , \beta = 1,\ldots , 4 \, .
\end{equation}
The subgroups $U(n)$ and $U(2,2)$ of $Mp(8n)$ form in this case a dual pair, the Weyl generators of $u(2,2)$,
\begin{equation}
\label{eq4.5}
\sum_{i=1}^n \tilde\varphi_{i\alpha} \, \varphi_i^{\beta}
\end{equation}
being $U(n)$ invariant. The Lagrangian subspaces ${\mathbb C}^n \otimes {\mathcal V}$ and ${\mathbb C}^n \otimes {\mathcal V}'$ (spanned by $\varphi_i^{\alpha}$ and $\tilde\varphi_{j\beta}$, respectively) are both $U(n)$ and $U(2,2)$ invariant, so that our dual pair is indeed of type II.

\smallskip

One of the elements of each dual pair ($O(m)$ for type I and $U(n)$ for type II) in the above examples is a compact group. This is, sure, not the general case but our choice is motivated physically: it allows to interpret this element as a global {\it gauge group}. We shall encounter infinite dimensional generalization of these examples in the quantum field theory context in Section~6.

\section{Global conformal invariance and infinite dimensional Lie algebras}\label{sec5}
\setcounter{equation}{0}

It has been often argued that 2D CFT methods are much too special to allow an extension to higher dimensions:

\smallskip

\noindent (i) the appearance of a chiral current algebra with meromorphic correlation functions is usually related to the splitting of the ($1+1$)-dimensional light-cone into a product of two light rays;

\smallskip

\noindent (ii) the representation theory of Kac-Moody and of Virasoro algebras is playing a crucial role in constructing soluble 2D CFT models; there are, on the other hand, no local Lie fields in higher dimensions (see \cite{B08} and \cite{T08} for a more detailed discussion and for references to the no-go theorem of Robinson and Baumann).

\smallskip

It was revealed, however, that the existence of a current algebra with rational correlation functions is due to a symmetry property: {\it global} (rather than infinitesimal) {\it conformal invariant} (GCI) that implies a strong (Huygens) locality condition and thereby rationality of correlation functions \cite{NT} in any even number of space-time dimensions. Indeed, the fact that global conformal transformations can map a space-like onto a time-like pair of events, together with local commutativity for space-like separations imply that local commutators are supported on the light cone
\begin{equation}
\label{eq5.1}
(x_{12}^2)^n \, [\phi_1 (x_1) , \phi_2 (x_2)] = 0 \quad \mbox{for} \quad x_{12}^2 := ({\bm x}_1 - {\bm x}_2)^2 - (x_1^0 - x_2^0)^2 , \ n \gg 0
\end{equation}
($n \gg 0$ meaning $n$ sufficiently large). Combined with the remaining Wightman axioms (in particular, with the energy positivity condition) this implies rationality of correlation functions of local GCI fields \cite{NT}; more precisely, $n$-point Wightman functions are proven to have the form
\begin{equation}
\label{eq5.2}
{\mathcal W}_n (x_1 , \ldots , x_n) = \frac{P(x_{jk})}{\underset{j < k}{\prod} \
 (x_{jk}^2)_+^{\mu_{jk}}} \, , \ \mu_{jk} \in {\mathbb N} \, , \ P \ \mbox{a polynomial,} \ (x^2)_+ = x^2 + i\epsilon x^0.
\end{equation}

Furthermore, although there are no local scalar Lie fields in four (or higher) 
dimensions (\cite{G68})(\cite{B76}), infinite dimensional Lie algebras are 
generated by {\it bifields} $V(x_1 , x_2)$ that naturally arise in the operator
 product expansions of a finite set of (say, hermitean, scalar) fields $\phi_i$
 of dimension $d$ ($>1$):
\begin{equation}
\label{eq5.3}
(x_{12}^2)^d \, \phi_i(x_1) \, \phi_j (x_2) = N_{ij} + x_{12}^2 \, V_{ij} (x_1 , x_2) + O((x_{12}^2)^2) \, ,
\end{equation}
where $V_{ij}$ are defined as infinite sums of contributions of (twist two) local conserved tensor currents (and the real symmetric matrix $N_{ij}$ is positive definite). These bifields are proven (as a consequence of the local conservation laws) to be harmonic in each variable (\cite{NST03}):
\begin{equation}
\label{eq5.4}
\Box_1 \, V_{ij} (x_1 , x_2) = 0 = \Box_2 \, V_{ij} (x_1 , x_2) \, , \ \Box_k = \frac{\partial}{\partial \, x_k^{\mu}} \, \frac{\partial}{\partial \, x_{k\mu}} \, .
\end{equation}
It can be deduced from the analysis of 4-point functions that the commutator algebra of the harmonic bifields $V_{ij}$ appearing in (\ref{eq5.1}) only closes on the  $V$'s and the unit operator if the local GCI scalar fields $\phi_j$ have dimension $d=2$ ($d=1$ corresponding to a free field theory). In that case the bifields $V$ are proven to be Huygens bilocal \cite{NRT}.

\smallskip

The infinite Lie algebras generated by such bilocal fields are characterized by the following result of \cite{NRT}.

\bigskip

\noindent {\bf Theorem 5.1.} -- {\it The harmonic bilocal fields $V$ arising in the expansion} (\ref{eq5.3}) {\it of the GCI hermitean scalar fields $\phi_j$ of dimension $d=2$ can be labeled by elements $M$ of a unital real matrix algebra ${\mathcal M} \subset {\rm Mat} \, (L,{\mathbb R})$ closed under transposition $M \to \ ^tM$ in such a way that the commutator of $V_M$ and $V_{M'}$ be given by
\begin{eqnarray}
\label{eq5.5}
&&[V_M (x_1 , x_2) , V_{M'} (x_3 , x_4)] = \Delta_{13} \, V_{^tMM'} (x_2 , x_4) + \nonumber \\
&+ &\Delta_{24} \, V_{M \, ^tM'} (x_1 , x_3) + \Delta_{23} \, V_{MM'} (x_1 , x_4) + \Delta_{14} \, V_{M'M} (x_3 , x_2) + \nonumber \\
&+ &{\rm tr} \, (MM') \, \Delta_{12,34} + {\rm tr} \, (^tMM') \, \Delta_{12,43} \, .
\end{eqnarray}
Here $\Delta_{ij}$ is the free field commutator, $\Delta_{ij} := \Delta_{ij}^+ - \Delta_{ji}^+$, and $\Delta_{12 , ij} = \Delta_{1i}^+ \, \Delta_{2j}^+ - \Delta_{i1}^+ \, \Delta_{j2}^+$ where
\begin{equation}
\label{eq5.6}
\Delta_{12}^+ = \int_{p_0 = \vert {\bm p} \vert} e^{-ipx_{12}} \, \frac{d^{3} p}{(2\pi)^3 \, 2p_0} = \frac{1}{4\pi^2 \, (x_{12}^2)_+} \, , \ x_{12} = x_1- x_2
\end{equation}
is the $2$-point Wightman function of a free massless scalar field.}

\bigskip

The matrix algebra ${\mathcal M}$ is an example of a (finite dimensional) {\it real star algebra}; we shall call it a $t$-{\it algebra} as the star operation is just matrix transposition. It can be equipped with a {\it Frobenius inner product}
\begin{equation}
\label{eq5.7}
\langle M_1 , M_2 \rangle = {\rm tr} \, (^tM_1 \, M_2) = \sum_{ij} M_{1ij} \, M_{2ij}
\end{equation}
which is symmetric, positive definite and has the property
\begin{equation}
\label{eq5.8}
\langle M_1 \, M_2 , M_3 \rangle = \langle M_1 , M_3 \ ^tM_2 \rangle \, .
\end{equation}
This implies that for every right ideal ${\mathcal I} \subset {\mathcal M}$ its orthogonal complement is again a right ideal while its transposed $^t{\mathcal I}$ is a left ideal. It follows that the algebra ${\mathcal M}$ is semi-simple so that every ${\mathcal M}$-module is a  direct sum of irreducible modules.

\medskip

Let now ${\mathcal M}$ be irreducible. It then follows from Schur's lemma (whose real version, \cite{L}, is more involved than the better known complex one) that its commutant ${\mathcal M}'$ in ${\rm Mat} \, (L,{\mathbb R})$ coincides with one of the {\it three real division rings}: the fields of real and complex numbers, ${\mathbb R}$ and ${\mathbb C}$, or the (noncommutative) ring ${\mathbb H}$ of quaternions. In each case the Lie algebra of bilocal fields is a central extension of an infinite dimensional Lie algebra that admits a discrete series of highest weight representations. Moreover, it was proven (first in the theory of a single scalar field of dimension 2 \cite{NST} and eventually for an arbitrary set of such fields \cite{NRT}) that in each case we are dealing with a {\it minimal representation} appearing as a subrepresentation of the Heisenberg algebra. More precisely, the bilocal field $V_M$ can always be written as a linear combination of normal products of free massless scalar fields $\varphi$. For each of the above irreducible types $V_M$ has a canonical form:
\begin{eqnarray}
\label{eq5.9}
&{\mathbb R} : &V(x_1 , x_2) = \sum_{i=1}^N : \varphi_i (x_1) \, \varphi_i (x_2): \nonumber \\
&{\mathbb C} : &W(x_1 , x_2) = \sum_{j=1}^N : \varphi_j^* (x_1) \, \varphi_j (x_2): \nonumber \\
&{\mathbb H} : &Y(x_1 , x_2) = \sum_{m=1}^N : \varphi_m^+ (x_1) \, \varphi_m (x_2): \, ,
\end{eqnarray}
where $\varphi_i$ are real, $\varphi_j$ are complex and $\varphi_m$ are quaternionic valued fields (for $L=N$, $2N$, and $4N$, respectively). We shall denote the corresponding infinite dimensional Lie algebra by ${\mathcal L} ({\mathbb F})$ for ${\mathbb F} = {\mathbb R}$, ${\mathbb C}$, or ${\mathbb H}$.

\smallskip

In order to identify the types of Lie algebras ${\mathcal L} ({\mathbb F})$ it 
is convenient to introduce a discrete basis. There is a natural expansion of the free massless field $\varphi$ into discrete modes since $\varphi$ can be viewed (as any GCI field) as defined on the conformal compactification $\bar M$ of Minkowski space $M$. To this end it is handy to use the complex variable parametrization
\begin{eqnarray}
\label{eq5.10}
\bar M &= &\left\{ z = (z_{\alpha} , \alpha = 1 , \ldots , 4); z_{\alpha} = \ell^{it} \, u_{\alpha} , t , u_{\alpha} \in {\mathbb R} , u^2 := \sum_{\alpha = 1}^4 u_{\alpha}^2 = 1 \right\} \nonumber \\
&= &{\mathbb S}^3 \times {\mathbb S}^1 / {\mathbb Z} / (2) \, ,
\end{eqnarray}
introduced in \cite{T86} (and later explored in \cite{N} and \cite{NT05}). $M$ is embedded in $\bar M$ by setting
\begin{equation}
\label{eq5.11}
{\bm z} = \frac{{\bm x}}{\omega (x)} \, ({\bm x} = (x^1 , x^2 , x^3)) \, , \ z_4 = \frac{1-x^2}{2 \, \omega (x)} \, , \ 2 \, \omega (x) = 1+x^2 - 2 \, i \, x^0
\end{equation}
$$
\left( x^2 := {\bm x}^2 - (x^0)^2 \, , \ {\bm x}^2 = \sum_{i=1}^3 (x^i)^2 \right) \, , \ \varphi_M (x) = \frac{1}{2\pi \omega (x)} \, \varphi_{\bar M} (z(x)) \, .
$$

\bigskip

\noindent {\bf Remark 5.1.} Factoring ${\mathbb Z} / (2) = \{ id , (t,u) \to (t+\pi , -u)\}$ in (\ref{eq5.10}) makes sense if we view $\bar M$ as a (real) pseudo-Riemannian manifold with metric
\begin{equation}
\label{eq5.12}
ds^2 = \frac{dz^2}{z^2} = du^2 - dt^2 \quad \left( \mbox{for} \ u \, du = \sum_{\alpha = 1}^4 u_{\alpha} \, du_{\alpha} = 0 \right)
\end{equation}
or if we equip it with a group structure, setting $\bar M = U(2)$ (which is only possible in four space-time dimensions -- cf.~\cite{U}). Viewed as just a smooth manifold ${\mathbb S}^3 \times {\mathbb S}^1 / {\mathbb Z} / (2)$ is diffeomorphic to ${\mathbb S}^3 \times {\mathbb S}^1$.

\smallskip

A free massless field $\varphi (z)$ on $\bar M$ admits an expansion of the form
\begin{equation}
\label{eq5.13}
\varphi (z) = \sum_{n=1}^{\infty} \ \sum_{\ell = 0}^{n-1} \ \sum_{m=-\ell}^{\ell} \left( \frac{a_{n\ell m}}{(z^2)^n} + a_{n\ell m}^* \right)  h_{n\ell m} (z) \, ,
\end{equation}
where $h_{n\ell m}$ are (suitably normalized) homogeneous (of degree $n-1$) harmonic polynomials, the {\it spherical harmonics}, satisfying
\begin{equation}
\label{eq5.14}
(H-n) \, h_{n\ell m} (z) = 0 \quad \mbox{for} \quad H = z \cdot \frac{\partial}{\partial \, z} + 1
\end{equation}
($H$ is the {\it conformal Hamiltonian} that generates translations in $t$),
\begin{equation}
\label{eq5.15}
{\bm L}^2 \, h_{n \ell m} (z) = \ell (\ell + 1) \, h_{n\ell m} (z) \, , \ (L_3 - m) \, h_{n\ell m} (z) = 0
\end{equation}
${\bm L} = (L_1 , L_2 , L_3)$ being the angular momentum vector operator,
\begin{equation}
\label{eq5.16}
L_j = i \, \varepsilon_{jk\ell} \, z_{\ell} \, \frac{\partial}{\partial \, z_k} \qquad \left( L_3 = i \left( z_2 \, \frac{\partial}{\partial \, z_1} - z_1 \, \frac{\partial}{\partial \, z_2} \right)\right) \, ,
\end{equation}
while $a_{\nu}^{(*)}$ (where $\nu$ is the triple index, $\nu = (n,\ell , m)$) obey the CCR
\begin{equation}
\label{eq5.17}
[a_{\nu} , a_{\nu'}^*] = \delta_{\nu\nu'} (= \delta_{nn'} \, \delta_{\ell\ell'} \, \delta_{mm'}) \, .
\end{equation}
The bilocal fields (\ref{eq5.9}) are then expanded into products of $a_{\nu}^{(*)} , a_{\nu'}^{(*)}$. Cutting this expansion up to some finite energy level $n$ we can then identify the CR (\ref{eq5.5}) with the CRs of a classical (real) Lie algebra.

\section{The dual pairs $({\mathcal L} ({\mathbb F}) , U ({\mathbb F},N))$ and their Fock space representations. The Lie algebra $so^* (4n)$}\label{sec6}
\setcounter{equation}{0}

The analysis of \cite{B07} and \cite{B08} yields the following

\bigskip

\noindent {\bf Proposition 6.1.} {\it The algebras ${\mathcal L} ({\mathbb F})$, ${\mathbb F} = {\mathbb R} , {\mathbb C}, {\mathbb H}$ are $1$-parameter central extensions of appropriate completions of the inductive limits of classical real matrix algebras:
\begin{eqnarray}
\label{eq6.1}
{\mathbb R} &: &sp(2\infty , {\mathbb R}) = \lim_{n \to \infty} sp (2n,{\mathbb R}) \nonumber \\
{\mathbb C} &: &u(\infty , \infty) = \lim_{n \to \infty} u(n,n) \nonumber \\
{\mathbb H} &: &so^* (4\infty) = \lim_{n \to \infty} so^* (4n) \, .
\end{eqnarray}
In the free field realization} (\ref{eq5.9}) {\it the suitably normalized central charge coincides with the positive integer $N$. Moreover, in that case the corresponding bilocal field $(V,W,Y)$ is left invariant under a transformation of the constituent fields $\varphi$ by the compact gauge group $U(N,{\mathbb F})$ where}
\begin{equation}
\label{eq6.2}
U(N,{\mathbb R}) = O(N) \, , \ U(N,{\mathbb C}) = U(N) \, , \ U(N,{\mathbb H}) = Sp (2N) \ (\equiv USp (2N)) \, .
\end{equation}

\bigskip

It is remarkable (albeit disappointing from a physicist's point of view) that the free field realization holds in general.

\bigskip

\noindent {\bf Theorem 6.2.} (i) {\it In any unitary irreducible positive energy representation (UIPER) of ${\mathcal L} ({\mathbb F})$ the central charge $N$ is a positive integer.}

\smallskip

\noindent (ii) {\it All UIPERs of ${\mathcal L} ({\mathbb F})$ are realized (with multiplicities) in the Fock space ${\mathcal F}$ of $N \times \dim_{{\mathbb R}}  {\mathbb F}$ free hermitean massless scalar fields.}

\smallskip

\noindent (iii) {\it The ground states of equivalent UIPERs in ${\mathcal F}$ form irreducible representations of the gauge group $U(N,{\mathbb F})$} (\ref{eq6.2}). {\it This establishes a one-to-one correspondence between UIPERs of ${\mathcal L} ({\mathbb F})$ in the Fock space and the (finite dimensional) irreducible representations of $U(N,{\mathbb F})$.}

\bigskip

The {\it proof} of this theorem for ${\mathbb F} = {\mathbb R} , {\mathbb C}$ is given in \cite{B07} (the proof of (i) for the case of a single scalar field $\phi$ is already contained in \cite{NST}); the proof for ${\mathbb F} = {\mathbb H}$ is given in \cite{B08}.

\smallskip

The Lie algebras $sp (2n,{\mathbb R})$ and $u(n,n)$ were introduced in Section~3. The Lie algebra $so^* (4n)$, associated with the quaternions, can be defined as a subalgebra of the algebra $u(2n,2n)$ of $4n \times 4n$ complex matrices,
\begin{equation}
\label{eq6.3}
u(2n,2n) = \left\{ X \in {\rm Mat} \, (4n,{\mathbb C}) ; X^* \beta = -\beta X , \ \mbox{for} \ \beta = \begin{pmatrix} \un &{\mathbb O} \\ {\mathbb O} &-\un \end{pmatrix} \right\}
\end{equation}
such that\footnote{As explained in \cite{BB} the resulting Lie algebra is isomorphic to the algebra of complex skew symmetric matrices $Y (\in so(4n,{\mathbb C}))$ such that $Y^* \begin{pmatrix} {\mathbb O} &\un \\ -\un &{\mathbb O} \end{pmatrix} = -  \begin{pmatrix} 0 &\un \\ -\un &0 \end{pmatrix} Y$. We share the feelings of Barut and Bracken who call this algebra {\it remarkable}.}
\begin{equation}
\label{eq6.4}
so^* (4n) = \left\{ X \in u(2n,2n) ; \ ^tX \sigma = - \sigma X \, , \ \sigma = \begin{pmatrix} {\mathbb O} &\un \\ \un &{\mathbb O} \end{pmatrix} \right\}
\end{equation}
($\un$ standing for the $2n \times 2n$ unit matrix). It consists of $4n \times 4n$ matrices that can be written in the following $2n \times 2n$ block form:
\begin{equation}
\label{eq6.5}
X = \begin{pmatrix} u &v \\ v^* &- \, ^tu \end{pmatrix} , \ u \in u(2n) \ (\mbox{i.e.} \ u+u^*=0) , \ v \in so (2n,{\mathbb C}) , \ \mbox{i.e.} \ v + \, ^tv = 0 \, .
\end{equation}
(In particular, ${\rm tr} \, X = 0$, so that $so^* (4n) \subset su(2n,2n)$.) We again view $u(2n,2n)$ (and hence $so^* (4n)$) as a sub Lie algebra of $sp(8n,{\mathbb R})$ and consider its oscillator representation in which the $4n \times 4n$ matrices $X$ (\ref{eq6.3}) are mapped into the Fock space operators
\begin{equation}
\label{eq6.6}
\tilde\varphi \, X \, \varphi \, , \ \varphi = \begin{pmatrix} a_1 \\ \ldots \\ a_{2n} \\ b_1^* \\ \ldots \\ b_{2n}^* \end{pmatrix} , \ \tilde\varphi = \varphi^* \beta = (a_1^* , \ldots , a_{2n}^* , -b_1 , \ldots , -b_{2n})
\end{equation}
(cf. Remark~3.1. This construction corresponds to $N=1$ in Proposition~6.1 and Theorem~6.2.) The Chevalley-Cartan basis of the subalgebra $so^* (4n)$ in this realization reads:
\begin{eqnarray}
\label{eq6.7}
E_i &= &a_i^* \, a_{i+1} + b_{i+1} \, b_i^* \, , \ F_i = E_i^* \, , \ i = 1,\ldots , 2n-1 \, , \nonumber \\
E_{2n} &= &a_{2n-1}^* \, b_{2n}^* - a_{2n}^* \, b_{2n-1}^* \, , \ F_{2n} = a_{2n} \, b_{2n-1} - a_{2n - 1} \, b_{2n} \, , \nonumber \\
H_i &= &a_i^* \, a_i - a_{i+1}^* \, a_{i+1} + b_i^* \, b_i - b_{i+1}^* \, b_{i+1} \, , \ i=1,\ldots , 2n-1 \, , \nonumber \\
H_{2n} &= &a_{2n-1}^* \, a_{2n-1} + a_{2n}^* \, a_{2n} + b_{2n-1} \, b_{2n-1}^* + b_{2n} \, b_{2n}^* \, .
\end{eqnarray}
The centre $u(1)$ of $u(2n,2n)$ is generated by the {\it charge operator}
\begin{equation}
\label{eq6.8}
Q = \sum_{i=1}^{2n} (a_i^* \, a_i - b_i^* \, b_i) =: a^* \cdot a - b^* \cdot b \, .
\end{equation}
The (non-compact) raising operator $E_{\theta}$ corresponding to the highest root $\theta = \alpha_{2n-2} + \underset{i=1}{\overset{2n}{\sum}} \, \alpha_i$ of $so^* (4n)$ is
\begin{equation}
\label{eq6.9}
E_{\theta} = E_{12} \, , \ \mbox{for} \ E_{ij} = a_i^* \, b_j^* - a_j^* \, b_i^* \, , \ 1 \leq i < j \leq 2n
\end{equation}
($E_{2n-12n} \equiv E_{2n}$). The commutant ${\mathcal H}$ of the principal $s\ell_2$ subalgebra of $so^* (4n)$ is $so^* (4n-4) \oplus su(2)$.

\bigskip

\noindent {\bf Proposition 6.3.} {\it The Lie algebras $sp(2) \simeq su(2)$ (of the compact group $Sp(2) (\equiv USp (2)) \simeq SU(2)$) with generators
\begin{equation}
\label{eq6.10}
E = a^* \cdot b \left( := \sum_{i=1}^{2n} a_i^* \, b_i \right) , \ F = b^* \cdot a \, , \ Q = [E,F] = a^* \, a - b^* \, b
\end{equation}
and $so^* (4n)$ form a dual pair in $sp(8n)$.}

\bigskip

This proposition enters as part in the argument (for a special case) of Theorem~6.2. Its proof uses the fact that the second order Casimir operator $C^{so}$ of $so^* (4n)$ is majorized by the Casimir $C^u$ of its maximal compact subalgebra $u(2n)$; more precisely, we have the relation
\begin{equation}
\label{eq6.11}
C^{so} = C^u - \sum_{1 \leq i < j \leq 2n} E_{ij} \, E_{ij}^* \quad \mbox{for} \quad C^u = C^{su} + \frac{H^2}{2n}
\end{equation}
where $C^{su}$ is the sandard quadratic Casimir operator of $su(2n)$ while
\begin{equation}
\label{eq6.12}
H = a^* \cdot a + b \cdot b^*
\end{equation}
is the generator of the centre $u(1)$ of $u (2n)$ (see \cite{B08}). Given the fundamental weights $\Lambda_i$ of $so^* (4n)$ (such that $(\Lambda_i \mid \alpha_j) = \delta_{ij}$ for $i,j = 1 , \ldots , 2n$, where $\alpha_j$ are the simple roots, $E_j = E_{a_j}$), the Fock space vacuum can be identified with $\vert 2 \, \Lambda_{2n} \rangle$:
\begin{equation}
\label{eq6.13}
(H_i - 2 \, \delta_{i2n}) \, \vert 2 \, \Lambda_{2n} \rangle = 0 = F_i \, \vert 2 \, \Lambda_{2n} \rangle = a_i \, \vert 2 \, \Lambda_{2n} \rangle = b_i \, \vert 2 \, \Lambda_{2n} \rangle \, .
\end{equation}
In accord with Proposition~6.3 (and Theorem~6.2) the vacuum is non-degenerate (it is an $SU(2)$ singlet), the UIPERs of lowest weight $2 \, \Lambda_{2n} - \Lambda_{2n-1}$ form an isospin doublet:
\begin{eqnarray}
\label{eq6.16}
F_i \, b_{2n}^* \, \vert 2 \, \Lambda_{2n} \rangle &= &0 = F_i \, a_{2n}^* \, \vert 2 \, \Lambda_{2n} \rangle \, , \quad i = 1,\ldots , 2n \, , \nonumber \\
E \, b_j^* \, \vert 2 \, \Lambda_{2n} \rangle &= &a_j^* \, \vert 2 \, \Lambda_{2n} \rangle \, , \quad j = 1,\ldots , 2n \, ;
\end{eqnarray}
$(a_{2n}^{*2} , a_{2n}^* \, b_{2n}^* , b_{2n}^{\times 2}) \, \vert 2 \, \Lambda_{2n} \rangle$ span an isotriplet of ground state vectors of the type $\vert 2 \, \Lambda_{2n} - 2 \, \Lambda_{2n-1} \rangle$ etc.

\bigskip

\noindent {\bf Remark 6.1.} For $n=2$ the group $so^* (8)$ is isomorphic to the conformal group in six dimensions \cite{BB},
\begin{equation}
\label{eq6.17}
so^* (8) \simeq so(6,2) \, .
\end{equation}
The degeneracy of the oscillator representation is reflected in the fact that the non-compact raising operators $E_{ij}$ satisfy second order equations of the type
\begin{equation}
\label{eq6.18}
E_{12} \, E_{34} + E_{14} \, E_{23} = E_{13} \, E_{24} \, .
\end{equation}
(For $so^* (8)$ this ``nilpotent cone relation'' is unique.)

\smallskip

Theorem 6.2 provides a link between two parallel developments, one in the study
 of highest weight modules of reductive Lie groups, the other in the work of
Doplicher-Haag-Roberts \cite{H} on the theory of global gauge symmetry and
superselection sectors (see \cite{B07} and \cite{T08} for a discussion and
references). The infinite Lie algebra ${\mathcal L} ({\mathbb F})$ and the
compact gauge group $U(N,{\mathbb F})$ appear as a rather special (limiting)
case of a reductive dual pair (Section~4). It would be interesting to explore
whether other (inequivalent) pairs are encountered in the commutator
algebra of (spin) tensor bifields that would naturally arise in the study of
(say, scalar) fields of higher (than two) scale dimension.

\bigskip

\noindent {\bf Acknowledgments.} It is a pleasure to thank my coauthors Bojko Bakalov, Nikolay M.~Nikolov and Karl-Henning Rehren together with whom are obtained the results reported in Sections~5 and 6.

\smallskip

I gratefully acknowledge the hospitality of SISSA, Trieste, IHES,
Bures-sur-Yvette and the Theory Unit of the Department of Physics of CERN
during the course of this work and the partial support of the Bulgarian
National Council for Scientific Research under contracts Ph-1406 and DO-02-257.

\end{document}